\documentclass[preprint,aps,graphicx]{revtex4}
\usepackage{graphicx}
\usepackage{dcolumn}
\usepackage{bm}

\begin{document}
\title{ Suppression of inelastic collisions of polar $^1 \Sigma$ state molecules in an electrostatic field}
\author{Alexander V.Avdeenkov$^{(1)}$  Masatoshi Kajita$^{(2)}$
and John L. Bohn$^{(3)}$} \affiliation{(1)Institute of Physics and
Power Engineering,
Obninsk,Kaluga region,249033, RUSSIA\\
(2)National Institute of Information and Communications Technology \\
4-2-1, Nukui-Kitamachi, Koganei, Tokyo 184-8795, JAPAN \\
(3) JILA, NIST and University of Colorado, Boulder, CO 80309 USA}
\date{\today}


\begin{abstract}
Collisions of polar $^{1}\Sigma$ state molecules at ultralow
energies are considered, within a model that accounts for
long-range dipole-dipole interactions, plus rotation of the
molecules.  We predict a substantial suppression of dipole-driven
inelastic collisions at high values of the applied electric field,
namely, field values of several times $B_e/\mu$. Here $B_e$ is the
rotational constant, and $\mu$ is the electric dipole moment of 
molecules.
 The sudden large drop in the
inelastic cross section is attributed to the onset of degeneracy
between molecular
rotational levels, which dramatically alters the scattering
Hamiltonian. As a result of the large ratio of elastic to inelastic
collision rates, we predict that evaporative cooling may be feasible
for $^{1}\Sigma$ state molecules in weak-field-seeking states,
provided a large bias electric field is present.
\end{abstract}
\maketitle

\section{Introduction}
The paramount goal of physics of ultracold temperatures is to
control and manipulate the quantum world. Polar molecules bring new
challenges and hopes to this field(see the review in \cite{Doyle0}).
 Since 1998~\cite{Doyle1}, several
experimental groups have been restraining polar molecules in order
to get colder and denser samples. The difficulties on this road
speak for themselves: a whole variety of novel experimental techniques were
developed for this purpose~\cite{Doyle0}. But in spite of  very intensive
experimental research, the production of ultracold molecules  still
poses a significant challenge.
 Cold and dense samples would allow one to control
 two- body~\cite{Balakrishnan,Bodo} and many-body systems of polar
particles~\cite{You,Goral1,Baranov1,
Baranov2,Baranov3,Yi1,Santos,Yi2,Goral2,Giovanazzi,Odell}, although so far
none of these goals has been experimentally realized. The main
obstacle to these achievements is loss of trapped molecules via inelastic
collisions, or else
Majorana transitions~\cite{Kajita0}. In
this paper we study just one aspect of physics of cold
polar molecules, namely, the possibility to have molecules in a
weak-field-seeking state with sufficiently low inelastic collision
rates to allow evaporative cooling.

We have previously considered the electrostatic trapping of polar
$\Pi$-state molecules of both bosonic~($OH$) and fermionic~($OD$)
symmetry from
the point of view  of stability with respect to collisions
\cite{Avdeenkov3,Avdeenkov4}. As  electro{\it static} trapping
requires molecules to be in a weak-electric-field-seeking state, collisions
involving the strong and anisotropic dipole-dipole interaction
between molecules may drive the molecules into unfavorable lower-energy
strong-field seeking states, leading to unacceptably high trap loss
and heating. For bosonic $\Pi$- state $OH$ molecules we have found
that the elastic rate can be much larger than the inelastic rate
only for quite large field values.
As the first excited rotational level of $OH$
lies $84K$ above the ground state, inelastic rates are defined
mostly by $\Lambda$- doubling
and a hyperfine splitting. In general one cannot yet exclude the
possibility of finding molecules whose hyperfine structure permits
evaporative cooling, but such a candidate has not yet been identified.

Polar fermions have a potentially important advantage for electrostatic
trapping, namely, low inelastic rates at cold temperatures. The
state-changing collisions of dipolar fermionic molecules were
discussed in \cite{Kajita1,Avdeenkov4}. Based on the well-known
Wigner threshold laws for dipole- dipole interactions it was shown
that elastic scattering cross sections are essentially independent
of collision energy $E$ at low energies, in electric fields
sufficiently strong to polarize the molecules~\cite{Avdeenkov4}).
At the same time,
state-changing cross sections scale as $E^{1/2}$ for fermions and
as $E^{-1/2}$ for bosonic molecules. Therefore, at ``sufficiently low''
temperatures, elastic scattering is always larger for fermions,
and evaporative cooling should be possible. Using the first
 Born approximation(BA), it was concluded~\cite{Kajita1}
that this is the case for the molecules OCS and CH$_3$Cl, at
reasonable experimental temperatures. However, the BA may not be strictly
applicable~\cite{Avdeenkov4} for all fields and energies of interest.
Indeed, the Fermi suppression of inelastic collisions may not be of great use
for evaporatively cooling the OD radical~\cite{Avdeenkov4}.

We are therefore motivated in this paper to re-visit the question
of field-dependent scattering of $^1\Sigma$ fermionic molecules, from the
perspective of close-coupling (CC) calculations.
A complete theoretical description of molecule-molecule scattering
is complicated by the complexity of the short-range interaction
between molecules. This interaction is generally unknown to sufficient
accuracy for cold collisions.
Therefore in order to avoid the inclusion of unknown parameters of
interaction, we seek and explore situations in which the influence of
short-range physics is minimal. It appears that for weak- field
seeking states the influence of the short-range potential is suppressed,
owing to avoided crossings in the long-range
interaction~\cite{Avdeenkov3}. For collisions of identical fermionic
molecules, the influence of short-range physics may be even smaller,
since only partial waves with $l \ge 1$ are present, and there is a
centrifugal repulsion in all scattering channels.

In this paper we seek a case where the elastic rate can be
significantly larger than the inelastic rate, by at least
two orders of magnitude, and so provide feasible
evaporative cooling \cite{Monroe}. The most important finding here is that a
large ratio of elastic and inelastic rates can be found
higher temperatures where the Wigner threshold law is not valid.
 Theoretically, we suggest that this phenomenon is quite general,
although large electric fields may be required.
 This effect exists even at temperatures
around $1mK$ which is already experimentally attained \cite{Doyle0}.
It is in just such a gas that evaporative cooling would be a useful
technique for achieving even higher phase-space density.

\section{Model}
\subsection{Polar $^1 \Sigma$- type molecules}

The majority of diatomic molecules  have $^{1}\Sigma$ electronic
ground states \cite{Herzberg}.  The energy  levels of these species can be
described by the rotation  $J$, total spin $F$ (i.e., including
nuclear spin), and vibration $\upsilon$
quantum numbers. In this paper for simplicity we will neglect
hyperfine splitting
as the hyperfine interaction for $^{1}\Sigma$ molecules is smaller
than for $\Pi $ or $%
^{3}\Sigma $ molecules and we consider them only in the $\upsilon =0
$ vibrational ground state. So we will treat polar molecules
as rigid rotors with a permanent dipole
moment. The Stark splitting will be characterized  by $(J,M_{J})$,
where $M_{J}$ is the  projection of $J$
on the direction of the external electric
field.  Thus the Hamiltonian for a polar $^1\Sigma$ molecule in a
field is
\begin{equation}
\label{hamone} H^{^1 \Sigma}=H_{rot}+H_{field}
\end{equation}
The matrix elements for the Hamiltonian (\ref{hamone}) in this basis
  are
\begin{eqnarray}
\label{matrix1} <J M_{J}| H^{^1 \Sigma}|J' M_{J'}>=
B_eJ(J+1)\delta_{JJ'}- \mu {\cal E} (-1)^{M_{J}}
 ([J][J'])^{1/2}
\left( \begin{array}{ccc}
                  J & 1 & J' \\
                  0 & 0 & 0
                  \end{array} \right)
\left( \begin{array}{ccc}
                  J & 1 & J' \\
                  -M_{J} & 0 & M_{J}
                  \end{array} \right)
\end{eqnarray}
In this expression $B_e$ is the rotational constant, $\mu$ is the
molecular dipole moment, ${\cal E}$ is the strength of the
electric field

The  different values of the molecular rotation $J$ are strongly
mixed in laboratory strength fields. Accordingly, in practice we
transform the molecular state to a field-dressed basis for
performing scattering calculations:
\begin{eqnarray}
\label{ebasis} |(\tilde{J})M_{J};{\cal E}
> \equiv \sum_{J} \alpha(J)
|JM_{J}>,
\end{eqnarray}
where $\alpha(J)$ stands for eigenfunctions of the
Hamiltonian~(\ref{hamone}) determined numerically at each value of
the field. The $J$ quantum number is not  a good quantum number
in a field, but we will continue to refer to molecular states with
$\tilde{J}$ as a reminder of the zero-field value of $J$.

Figure 1 shows the Stark energies computed using all the ingredients
described above. In zero field the energy levels are
determined by the rotational constant $B_e$. We demonstrate the Stark
splitting for linear OCS molecule which is quite typical for this
type of molecule. For the other molecules we consider, the figure
would look exactly the same, but with rescaled axes.
We estimated that the
hyperfine splitting is of order of $\mu K$, which means that this
effect may be important for ultracold energies. But here we will
ignore the hyperfine effects as the region of energies we are
considering is quite above  $\mu K$.  The Stark shift is quadratic for
fields below the ``critical field'' defined by ${\cal E}_{0}\equiv
B_eJ(J+1)/{2\mu}$.  It is a rather
approximate estimate because  of the mixing between the
neighboring rotation levels. For a molecule in its lowest weak-
field -seeking state $|10>$ the critical electric field is typically
on the order of $10^3 - 10^4$ V/cm for the species we consider here.
For fields larger than this,
the states with $J=0,1,2$ are deeply mixed. As a consequence, a
state like $|10 \rangle$, which is weak-field seeking in low fields,
can become high-field seeking at somewhat higher fields.  In the
following, as we are interested in $J=1$ states, the critical
field is given by ${\cal E}_{0}=B_e/\mu$.

\subsection{Dipole-dipole interaction}
The intermolecular dipole- dipole interaction has the form:
\begin{eqnarray}
\label{Dipole_Ham} V_{\mu \mu}({\bf R, \omega_1, \omega_2}) &=& {
\bm{\mu}_1 \cdot \bm{\mu}_2 - 3(\bm{{\hat R}}\cdot
\bm{\mu}_1)(\bm{{\hat R}}\cdot \bm{\mu}_2)
\over R^3}    \nonumber \\
&=& - {\sqrt{6} \over R^3} \sum_q (-1)^q C^{2}_{-q}(\omega) \left[
\mu_1 \otimes \mu_2 \right]^2_q.
\end{eqnarray}
where $\omega_{1,2}=(\theta_{1,2}, \phi_{1,2})$ are the polar
angles of molecules 1 and 2 with respect to the lab-fixed
quantization axis, and ${\bf R}=(R,\omega)$ is the vector between
the center of mass of the molecules in the laboratory fixed
coordinate frame. Here $C^{2}_{-q}(\omega)$ is a reduced spherical
harmonic~\cite{BS}.

We express the Hamiltonian in a basis of projection of total
angular momentum,
\begin{eqnarray}
\label{wave} {\cal M}= M_{J_{1}}+M_{J_{2}}+M_{l};
\end{eqnarray}
 $M_{l}$ is the
projection of the partial wave quantum number $l$ on the
laboratory axis. In this basis the wave function for two molecules
is described  as:
\begin{eqnarray}
\label{basis} \Psi^{\cal M}={1 \over R}\sum_{1,2,l,M_{l}} \left\{|1> \otimes
|2> \otimes |l M_{l}> \right\}^{\cal M} \times \psi^{{\cal
M},1,2}(R),
\end{eqnarray}
where $\left\{...\right\}^{\cal M}$ is the angular momentum  part of
this wave function and $|i>$ is the wave function for each molecule.
As we consider the target and the projectile as identical molecules,
we must take into account the symmetry of the wave
function~(\ref{basis}) under exchange.

Taking into account the Wigner-Eckart theorem, we can present the
reduced angular matrix element as
\begin{eqnarray}
\label{matrix2} <12lM_{l}|| A_{\Lambda}||1'2'l'M_{l'}>=
\\
\nonumber (-1)^{M_{J_{1}}' + M_{J_{2}}'+M_{l}-1}
([l][l'][J_{1}][J_{1}'][J_{2}][J_{2}'])^{1/2} \left(
\begin{array}{ccc}
                  1 & 1 & 2 \\
                  M_{J_{1}}-M_{J_{1}'} &  M_{J_{2}}-M_{J_{2}'} &  M_{l}-M_{l'}
                  \end{array} \right)
\times
\\
\nonumber \left( \begin{array}{ccc}
                  J_{1}' & 1 & J_{1} \\
                  0 &  0 &  0
                  \end{array} \right)
\left( \begin{array}{ccc}
                  J_{2}' & 1 & J_{2} \\
                  0 &  0 &  0
                  \end{array} \right)
\left( \begin{array}{ccc}
                  1 & J_{1} & J_{1}' \\
                  M_{J_{1}}-M_{J_{1}'} & -M_{J_{1}}&  M_{J_{1}'}
                  \end{array} \right)
\times
\\
\nonumber \left( \begin{array}{ccc}
                  1 & J_{2} & J_{2}' \\
                  M_{J_{2}}-M_{J_{2}'} & -M_{J_{2}}&  M_{J_{2}'}
                  \end{array} \right)
\left( \begin{array}{ccc}
                  l' & L & l \\
                  M_{l'} & M_{l}-M_{l'} &  -M_{l}
                  \end{array} \right)
\left( \begin{array}{ccc}
                  l' & 2 & l \\
                  0 & 0 &  0
                  \end{array} \right)
\end{eqnarray}
In practice, before each scattering calculation  the Hamiltonian
matrix   has to be transformed from this basis into the
field-dressed basis defined by (\ref{ebasis}). We solve the coupled-
channel equations using a logarithmic derivative propagator method
\cite{Johnson} to calculate total state-to-state cross sections.
Since the projection of total angular momentum on the field axis,
${\cal M}$, is a conserved quantity, calculations can be performed
for each value of ${\cal M}$ separately. We find, generally, that
the dominant contribution to cross sections arises from the minimum
allowed value of ${\cal M}$ and that the general behavior of cross
sections for other ${\cal M}$ is quite similar, and so restrict
calculations accordingly.

The scattering calculations quickly become computationally expensive
as more rotational states and partial waves are included.  Because
we need to calculate cross sections at many electric field values,
we choose a ``compromise'' basis set that includes rotational levels
up to $J_{max}=3$ and partial waves up to $L_{max}=3$. This
basis set then consists of 182 scattering channels.   This level
of approximation tends to get the general magnitude of elastic scattering
cross sections fairly accurately, and to {\it over}estimate
inelastic scattering cross sections.  We therefore expect to draw
conservative conclusions on the high ratio of elastic to inelastic
scattering.

\section{Results and discussion}

\begin{figure}
\centerline{\includegraphics[width=1.08\linewidth,height=0.9\linewidth,angle=0]{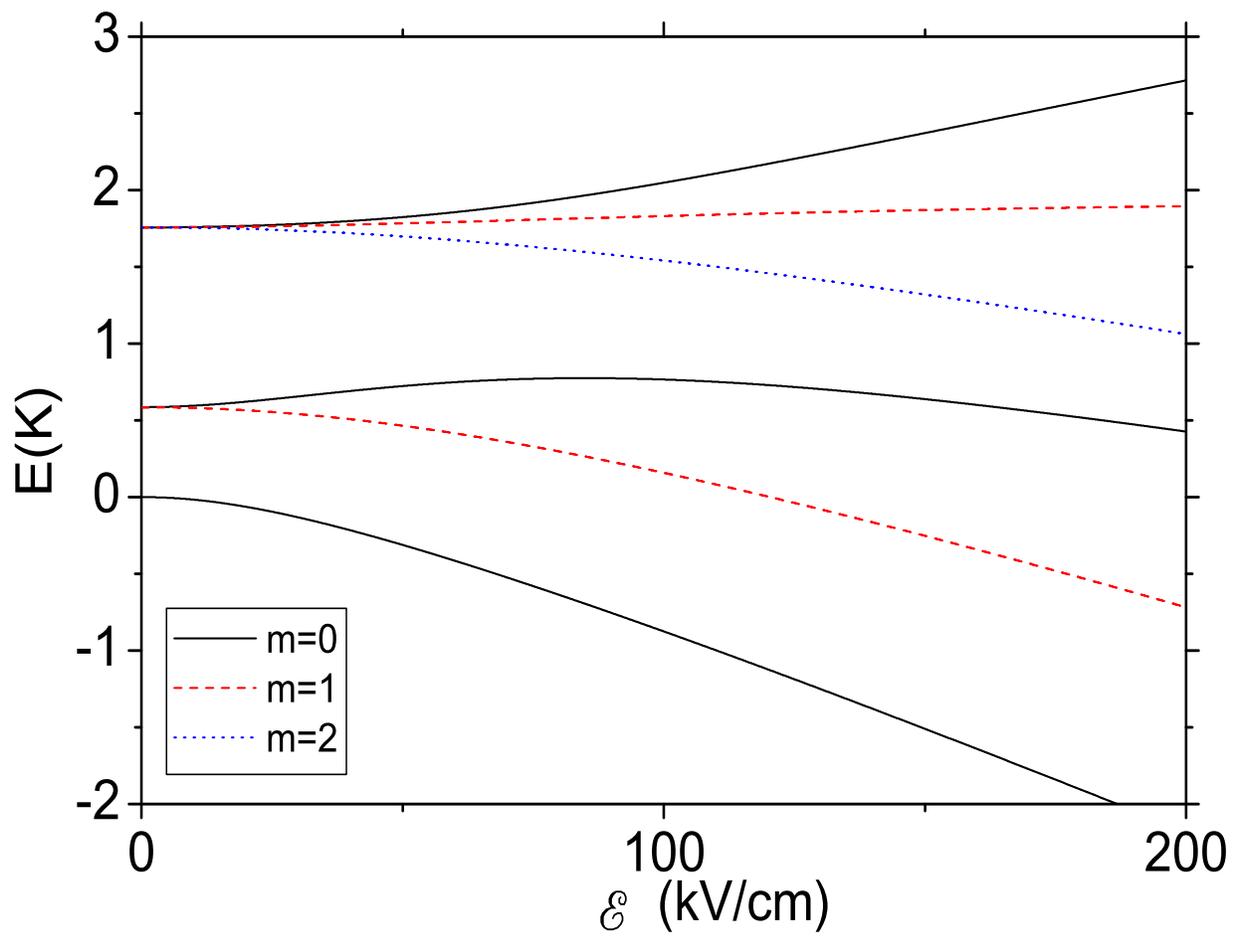}}
\caption{Stark effect for linear OCS molecules in their $^1
\Sigma$ ground state. } \label{stark}
\end{figure}

We have chosen a variety of different linear molecules for this study,
to span a range of rotational constants and dipole moments, and also
to connect with the results of Ref.~\cite{Kajita1}.  Properties
of the molecules are summarized in Table I~\cite{nist,RbCs}.
Of particular interest is the alkali dimer RbCs, which is a leading
candidate in the experimental quest to observe cold collisions
\cite{Sage}.  For these molecules, we are
interested in the lowest energy weak-field-seeking state of the
ground vibrational state, $| JM_{J}>=|10>$ (see Fig \ref{stark}).
In contrast to the $\Pi$ molecules we have previously studied
\cite{Avdeenkov1,Avdeenkov2}, here it is necessary to include several
rotational levels of the molecule.

\begin{figure}
\centerline{\includegraphics[width=1.08\linewidth,height=0.9\linewidth,angle=0]{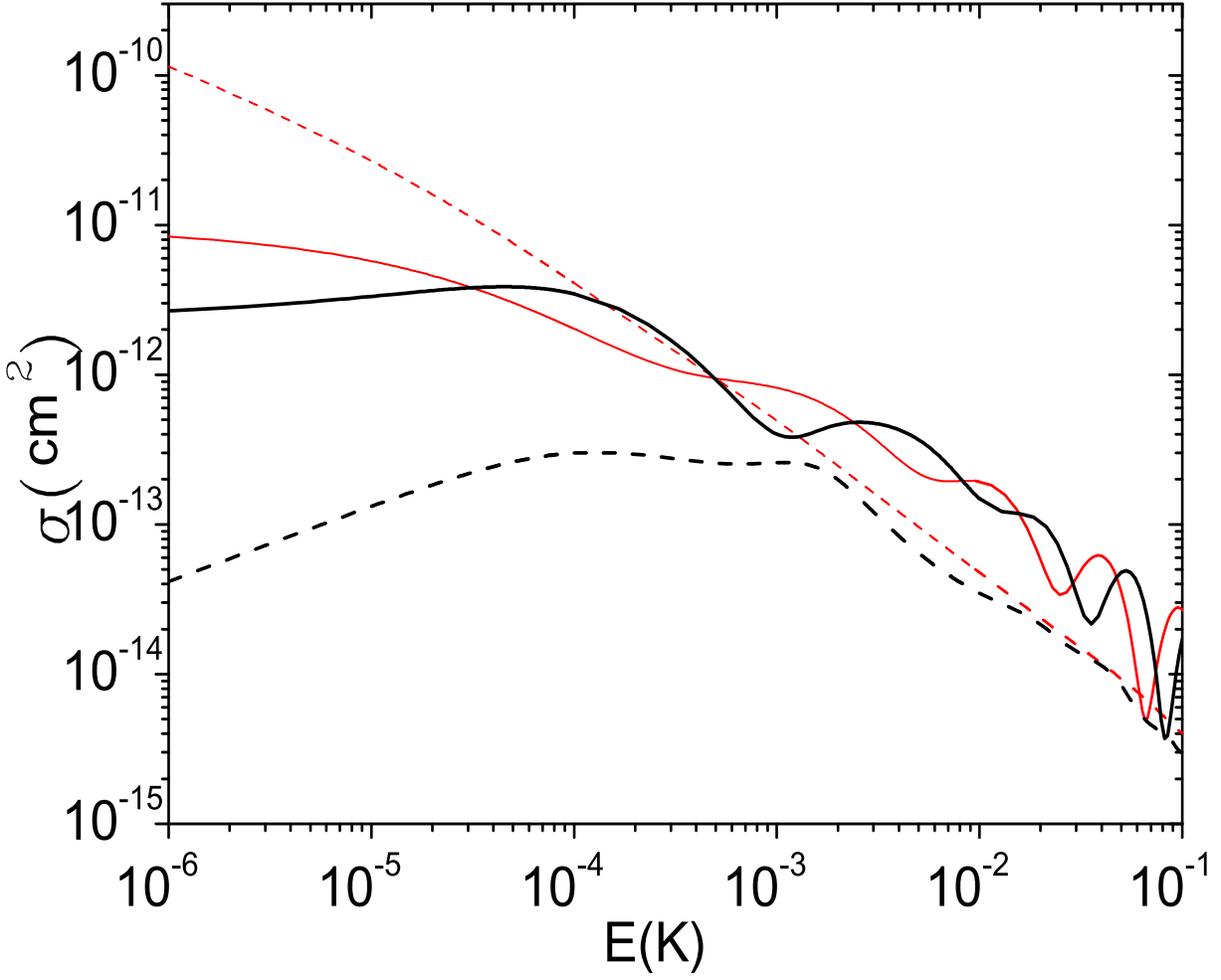}}
\caption{ Elastic(solid lines) and inelastic(dashed lines) cross
sections for ClCN molecule at an electrostatic field ${\cal E}=20000V/cm$.
Thick and thin curves are for Fermi and Bose particles
respectively.} \label{thresholds}
\end{figure}

The general behavior of the thermally-averaged cross sections
versus temperature is shown in Fig. \ref{thresholds}. This example
is for ClCN molecules in an electric field of ${\cal E}=20$ kV/cm,
well above the critical field for this molecule, meaning that the
molecule is strongly polarized. The heavy solid and dashed lines
represent (respectively) the elastic and state-changing cross
sections for fermionic isotopomers of  this molecule.  It can be
seen that the standard threshold behavior of colliding dipoles
occurs at energies below several tens of microKelvin.  Namely, the
elastic cross section becomes a constant, and the inelastic cross
section goes to zero as $T^{1/2}$ \cite{Avdeenkov4,Kajita1}.  By
contrast, the light solid and dashed curves show the same
quantities, but for a bosonic isotopomer.  Here the threshold laws
work against the experimentalist, with the inelastic rate
diverging as $T^{-1/2}$ \cite{Avdeenkov3,Kajita3}. At higher
temperatures, above the threshold regime, the elastic and
inelastic cross sections are comparable for both bosons and
fermions, consistent with what was found for OH and OD radicals in
Ref.\cite{Avdeenkov4}.

Remarkably, this situation can change quite dramatically if the
electric field takes certain values.  We illustrate this in Fig.
\ref{e_all}, by plotting elastic and inelastic cross sections at
$T=1$ $\mu$K, versus electric field.  In each case, the inelastic
cross sections (dashed lines) fall by 2-4 orders of magnitude at a
specific value of the electric field.  Here we show only results
for fermionic species. For bosonic species the corresponding
drop is smaller
because of the  threshold behavior at this energy.  The field
values at which this suppression occurs, denoted ${\cal E}_{\rm
supp}$, are tabulated in Table I.  In all cases the suppression
occurs at around 3 times the critical field ${\cal E}_0$.
To explore this phenomenon further, we focus on a particular
example in the following, namely, a fermionic isotopomer of
ClCN.

The unexpected suppression of inelastic collision rates has its
origin in the interplay between the strong dipole-dipole
interaction and the rotational energy levels of the molecules.
The first clue to the mechanism of suppression comes from
considering the scattering thresholds, i.e., the total energy of
both molecules when they are far apart. (Fig. \ref{thresh_E}).
Here each threshold is labelled according to the quantum numbers
$|J_1M_1, J_2M_2 \rangle$ of the colliding molecules, and zero
energy represents the lowest energy threshold corresponding to
$|00,00 \rangle$.  The solid line denotes the $|10,10 \rangle$
incident channel of interest to this paper.  It can be seen here
that this threshold crosses three other thresholds, for the
channels $|00,22\rangle$, $|00,21 \rangle$ and $|00,20 \rangle$,
at  fields ${\cal E}=11.6,12.8,13.65$ kV/cm respectively, very
near to ${\cal E}_{\rm supp}$ As we will see below ${\cal E}_{\rm
supp}$ can be defined more quantitatively as the field at which
thresholds for the channels $|10,10 \rangle$ and $|00,20 \rangle$
are degenerate.

This is a strange situation, in which the state-changing collision
cross sections {\it diminish} sharply just as new states become
energetically available.  The second piece to this puzzle is found
by examining the approximate adiabatic potential energy curves
for two situations, as shown in Fig.  \ref{adiab}.  To simplify
these curves, we have included only the partial wave $L=1$
in their construction, although $L=3$ is also used
in the multichannel scattering calculations.
In (a) are shown the curves for ${\cal E}=12.3$
kV/cm, just below the threshold crossing.  Here the incident
channel $|10,10 \rangle$ is below the nearby thresholds, and
correlates adiabatically to the solid curve.  This curve draws
molecules into the small-$R$ region where they interact
strongly and can readily change their internal state.

By contrast, Fig. \ref{adiab}(b) shows the adiabatic curves at a
field ${\cal E}=13.7$ kV/cm, just above the threshold crossing. Now
the incident channel correlates adiabatically to a repulsive curve,
so that the molecules do not approach one another nearly as closely
as in the previous case.  This shielding, in turn, reduces the
likelihood of inelastic collisions.  The dominance of the repulsive
curve is a peculiarity of the strong, anisotropic dipole-dipole
interaction, and is similar to the physics that generates
``field-linked'' molecular dimer states, which also keep the
molecules far from each other \cite{Avdeenkov1}. While this simple
picture of the suppression is probably a good first approximation,
there is clearly more going on. This can be seen in the tabulated
values of ${\cal E}_{\rm supp}$, which are not always equal to
$3{\cal E}_0$, even though this is the field value where the
thresholds cross for any $^1 \Sigma$ molecule.

We also remark that we have noted a similar, but somewhat less
dramatic, suppression of inelastic cross sections in bosonic
analogues of the molecules considered (Fig. \ref{e_bf}).
Presumably fermions have an
advantage since they have nonzero partial wave angular momentum in
all channels, which aids in the shielding effect discussed
above.

The strong suppression in the low-energy limit naturally has
consequences at higher collision energies.  Figure \ref{converg}
shows the elastic (solid) and inelastic (dashed) cross sections
versus temperature, at a field ${\cal E}=14$ kV/cm, where the
inelastic rates have just become suppressed.
 Strikingly, the ratio of
elastic to inelastic cross sections is close to two orders of
magnitude even at temperatures as high as several mK, which is
easily attainable in Stark deceleration or buffer-gas cooling
experiments.  For this reason, it is conceivable that cold, dense
samples might be amenable to evaporative cooling that will reduce
them to ultracold temperatures.

Figure \ref{converg} also illustrates the effect of increasing the
number of channels in the scattering calculations.  For both
elastic (solid lines) and inelastic (dashed lines) cross sections,
four curves are shown, corresponding to various maximal numbers of
rotational states ($J_{max}$) and partial waves ($L_{max}$).  In
all cases, changing the size of the basis set has little influence
on the overall elastic scattering cross section, although the
features understandably shift in field.  By contrast, The
inelastic cross sections are quite sensitive to $J_{max}$,
dropping more than an order of magnitude as $J_{max}$ increases
from 3 to 4. In addition, the number of partial waves plays a role
in the actual cross sections.  In any event, the conclusions still
hold, and the ratio of elastic to inelastic cross sections should
be quite high.

\section{Conclusion}
In this paper we considered the collisional dynamics of polar $^1
\Sigma$ molecules in a dc-electric field taking ClCN, HCN,
OCS and RbCs as prototypes. The lowest weak-field-seeking state was
studied from the point of view of evaporative cooling. As a
rule the strong and anisotropic dipole-dipole interaction  should
provide quite large inelastic rates. We have found, however,
 that polar $^1 \Sigma$
state molecules possess a substantial suppression of inelastic
collisions at high values of the applied electric field. The sudden
drop in inelastic cross section coincides with the degeneracy of
certain molecular rotational levels. Adiabatic pictures reveal that the
interaction changes from mostly attractive to mostly repulsive
upon crossing the field where this coincidence occurs.
The strong suppression of inelastic scattering from the $|JM_J \rangle=$
$|10 \rangle$ state of a $^1\Sigma$ molecule seems generally to occur
at a field nearly equal to $3B_e/\mu$, and may enable evaporative
cooling in an electrostatic trap.

As a final remark, we note that in many experiments alkali dimer molecules
are formed in high-lying vibrationally excited states, as a result
of photoassociation or magneto-association.  In these quite different states,
the field scale can be significantly larger.  To make an estimate,
consider the RbK molecule, which is being pursued by the UConn
group \cite{Wang}. In a hypothetical weakly-bound state whose
outer turning point is 40 a.u., this molecule's rotational constant
is approximately $B_e=1.4 $ cm$^{-1}$.  The dipole moment of this
molecule has been estimated to be approximately $1.4 \times 10^{-5}$
$ea_0$ at this intermolecular separation \cite{Kotochigova},
yielding an critical field of
${\cal E}_0 \approx 2 $ MV/cm.  Thus it would seem unlikely that
the effects described here are observable for such weakly-bound
molecules.  For even more weakly-bound molecules, the critical
field quickly becomes larger, owing to the exponential falloff
of the dipole moment.

This work was supported by the NSF and by a grant from the W. M. Keck
Foundation.


\newpage
\begin{table}
\caption{Molecular parameters for the species considered.  The rotational
constants $B_e$ and dipole moments $\mu$ for the triatomics come from
Ref. [28], while those for RbCs come from Ref. [29].  Here
${\cal E}_0= B_e/\mu$ is the ``critical'' field, while ${\cal E}_{\rm supp}$
is the calculated field value at which the inelastic processes become
suppressed. }
\begin{tabular}{lllll}
Molecule & $B_e$ (cm$^{-1}$) & $\mu$ (D) & ${3 \cal E}_0$ (kV/cm) & ${\cal E}_{\rm supp}$ \\
\hline
RbCs & 0.017 & 1.26 & 2.35 & 2.55 \\
ClCN & 0.199 & 2.833 & 12.57 & 13.65 \\
OCS & 0.203 & 0.715 & 50.71 & 55.3 \\
HCN & 1.478 & 2.985 & 88.51 & 96.4 \\
\end{tabular}
\end{table}

\begin{figure}
\centerline{\includegraphics[width=0.9\linewidth,height=0.65\linewidth,angle=0]{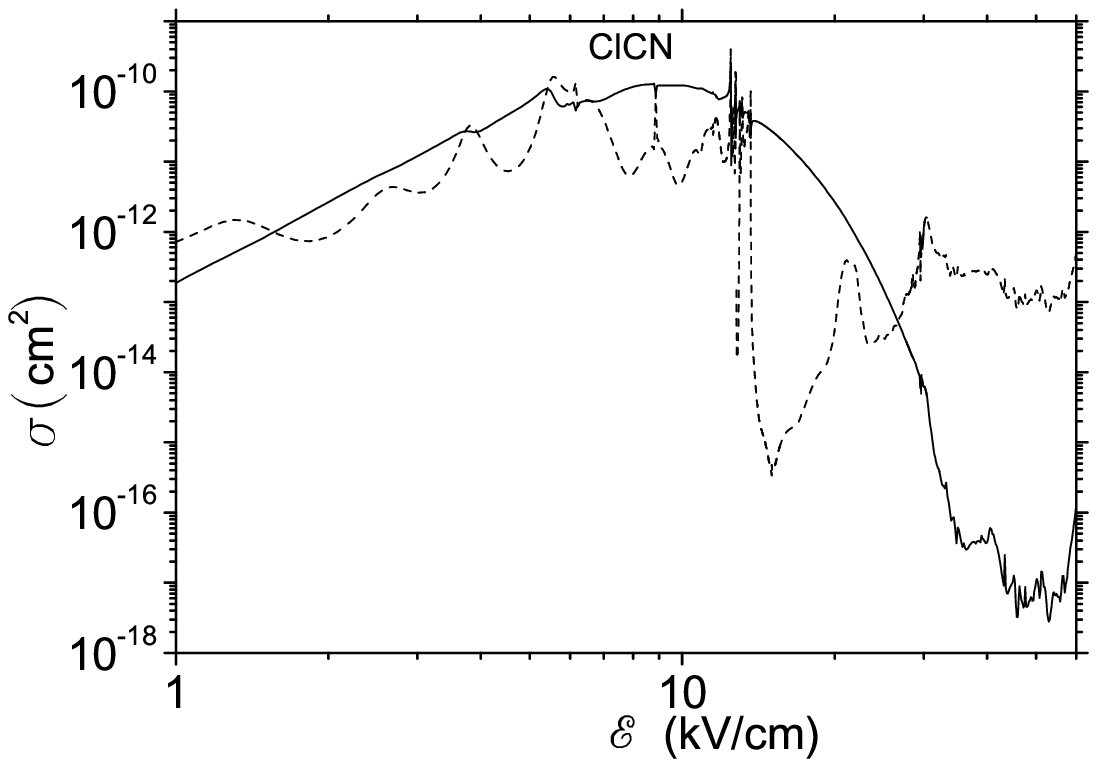}}
\centerline{\includegraphics[width=0.9\linewidth,height=0.65\linewidth,angle=0]{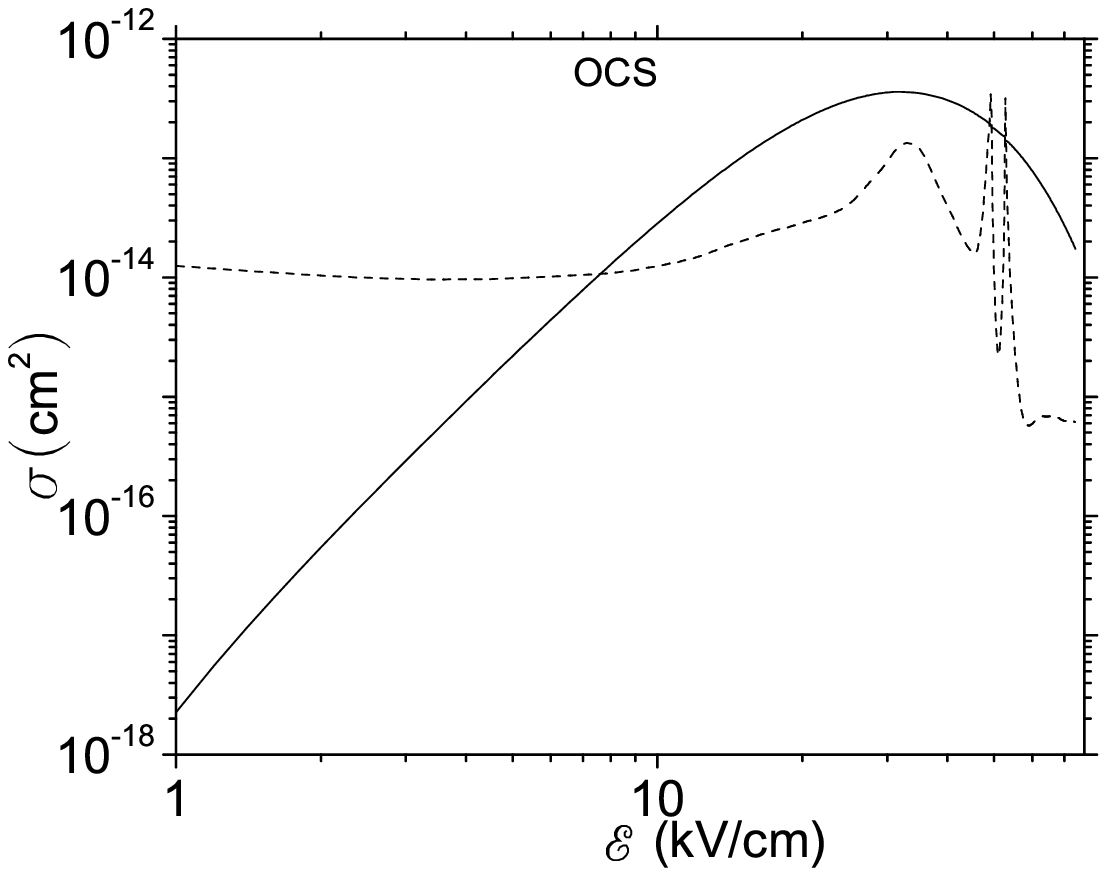}}
\end{figure}
\begin{figure}
\centerline{\includegraphics[width=0.9\linewidth,height=0.65\linewidth,angle=0]{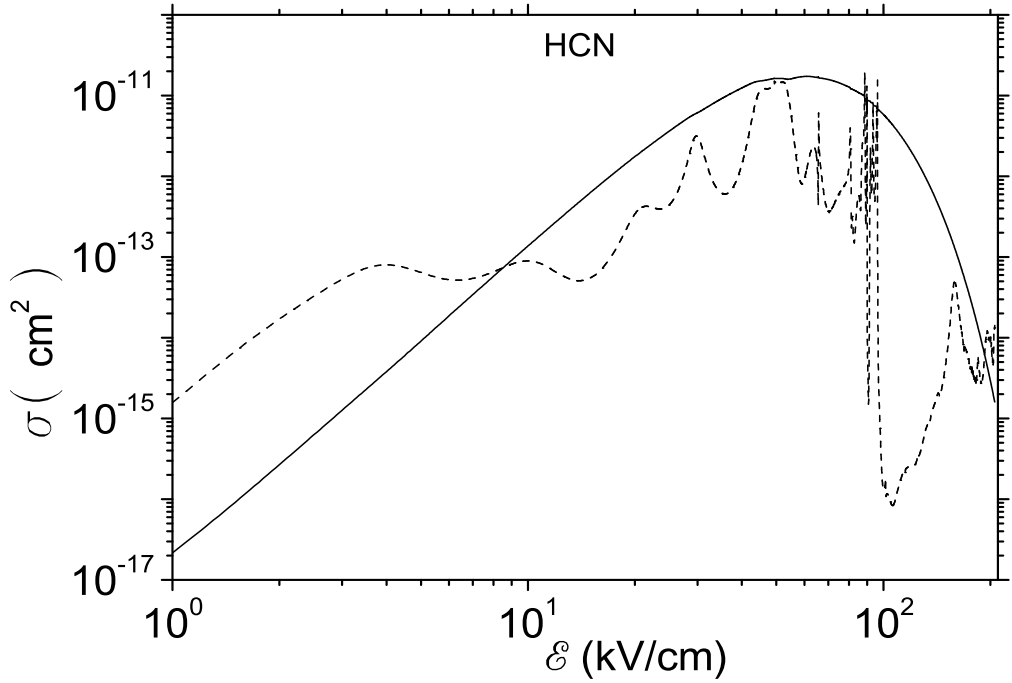}}
\centerline{\includegraphics[width=0.9\linewidth,height=0.65\linewidth,angle=0]{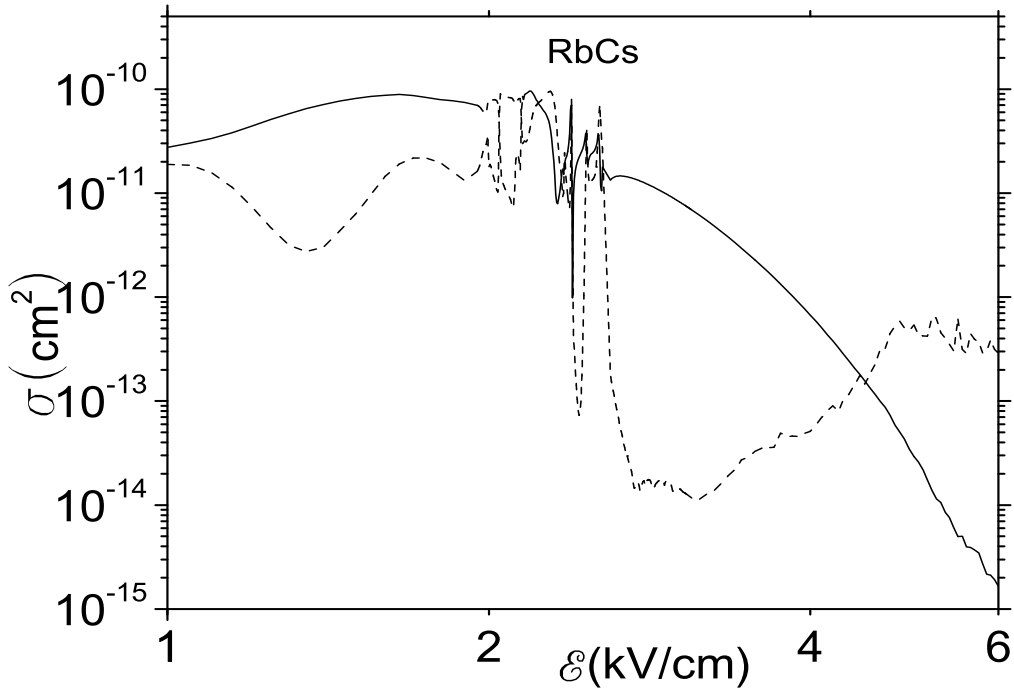}}
\caption{ Elastic(solid lines) and inelastic(dashed lines) cross
sections versus an electrostatic field at collision energy $1\mu
K$ for a)ClCN, b)OCS, c)HCN  and d)RbCs molecules. For these calculations,
$L_{max}=3, J_{max}=3$ }\label{e_all}
\end{figure}

\begin{figure}
\centerline{\includegraphics[width=1.08\linewidth,height=0.9\linewidth,angle=0]{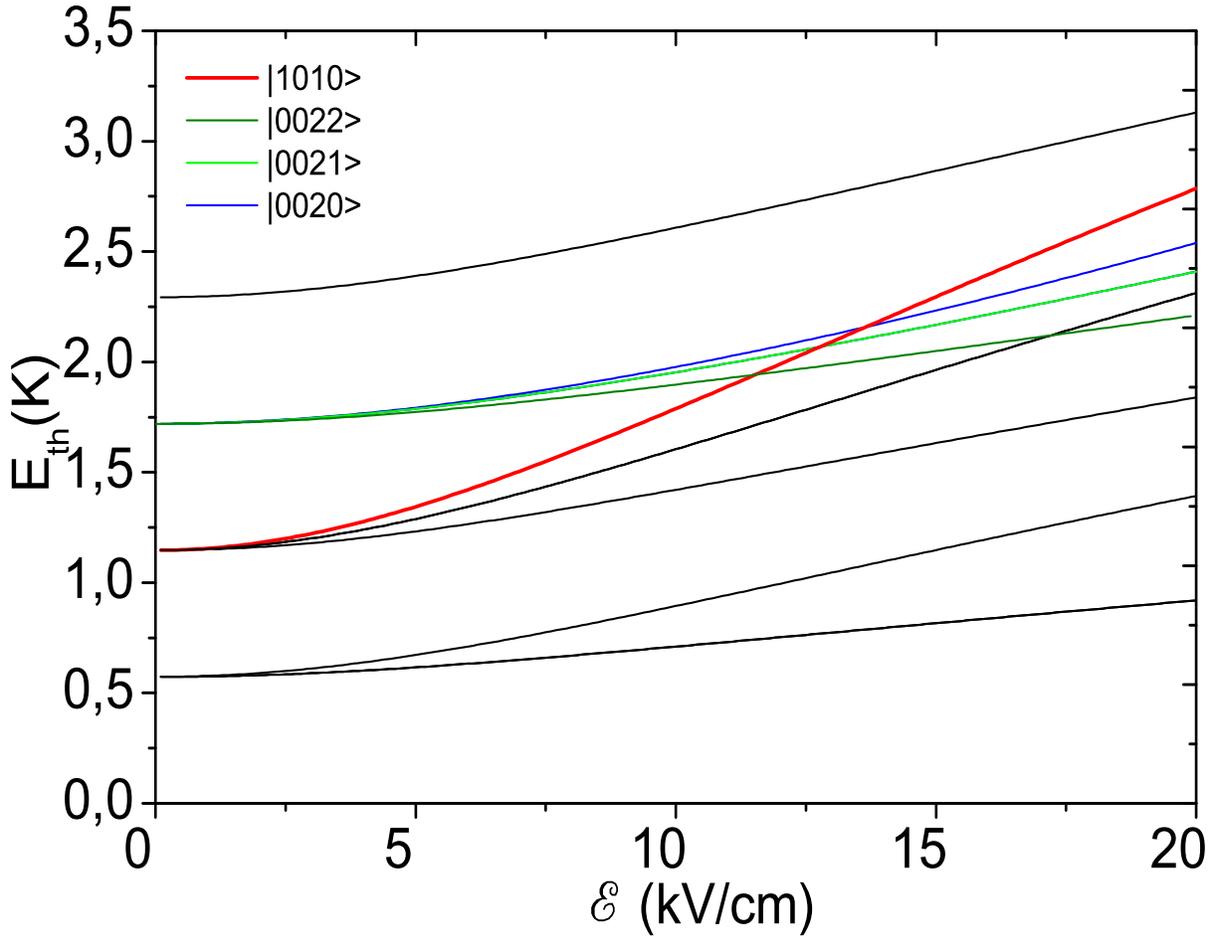}}
 \caption{Threshold energies for  ClCN molecules referred to
 the threshold of the $|00,00>$ channel. } \label{thresh_E}
\end{figure}

\begin{figure}
\centerline{\includegraphics[width=0.8\linewidth,height=0.6\linewidth,angle=0]{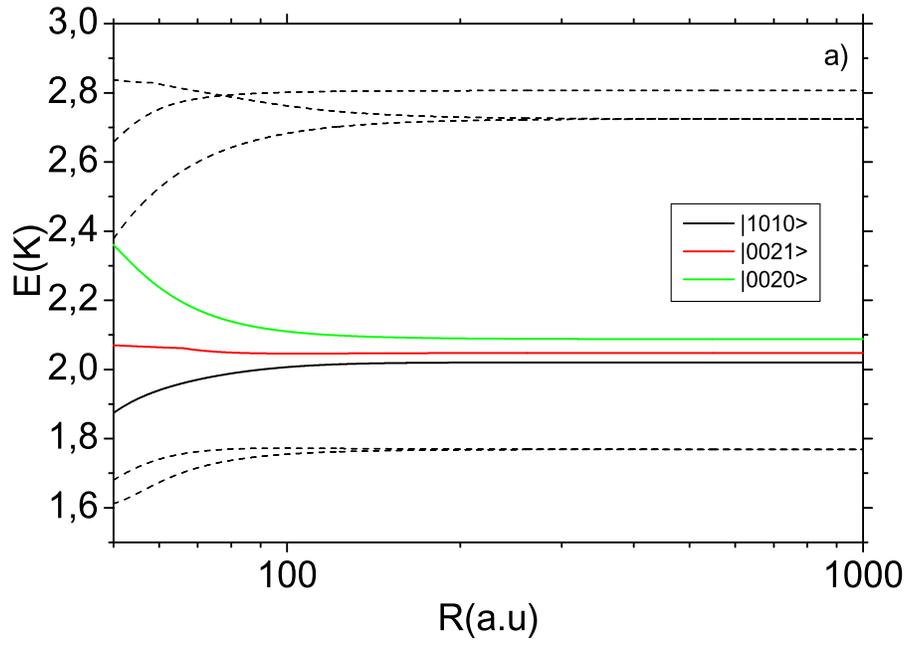}}
\centerline{\includegraphics[width=0.8\linewidth,height=0.6\linewidth,angle=0]{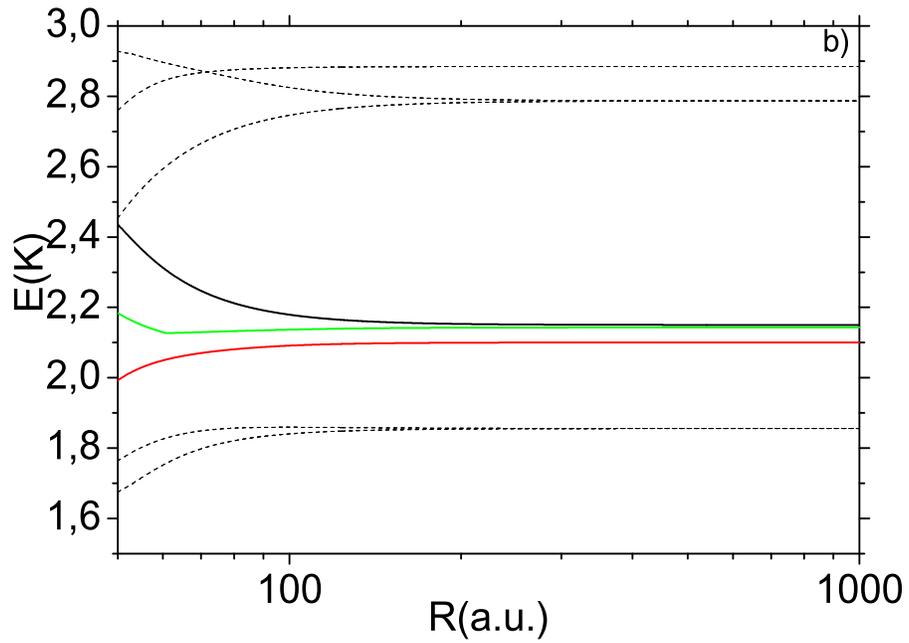}}
\caption{Adiabatic curves for ClCN molecule at a)12.3 kV/cm and
b)13.7 kV/cm.  In both cases, the dark curve correlates adiabatically
to the $|10\rangle |10 \rangle$ incident channel at large $R$.}
\label{adiab}
\end{figure}

\begin{figure}
\centerline{\includegraphics[width=1.08\linewidth,height=0.9\linewidth,angle=0]{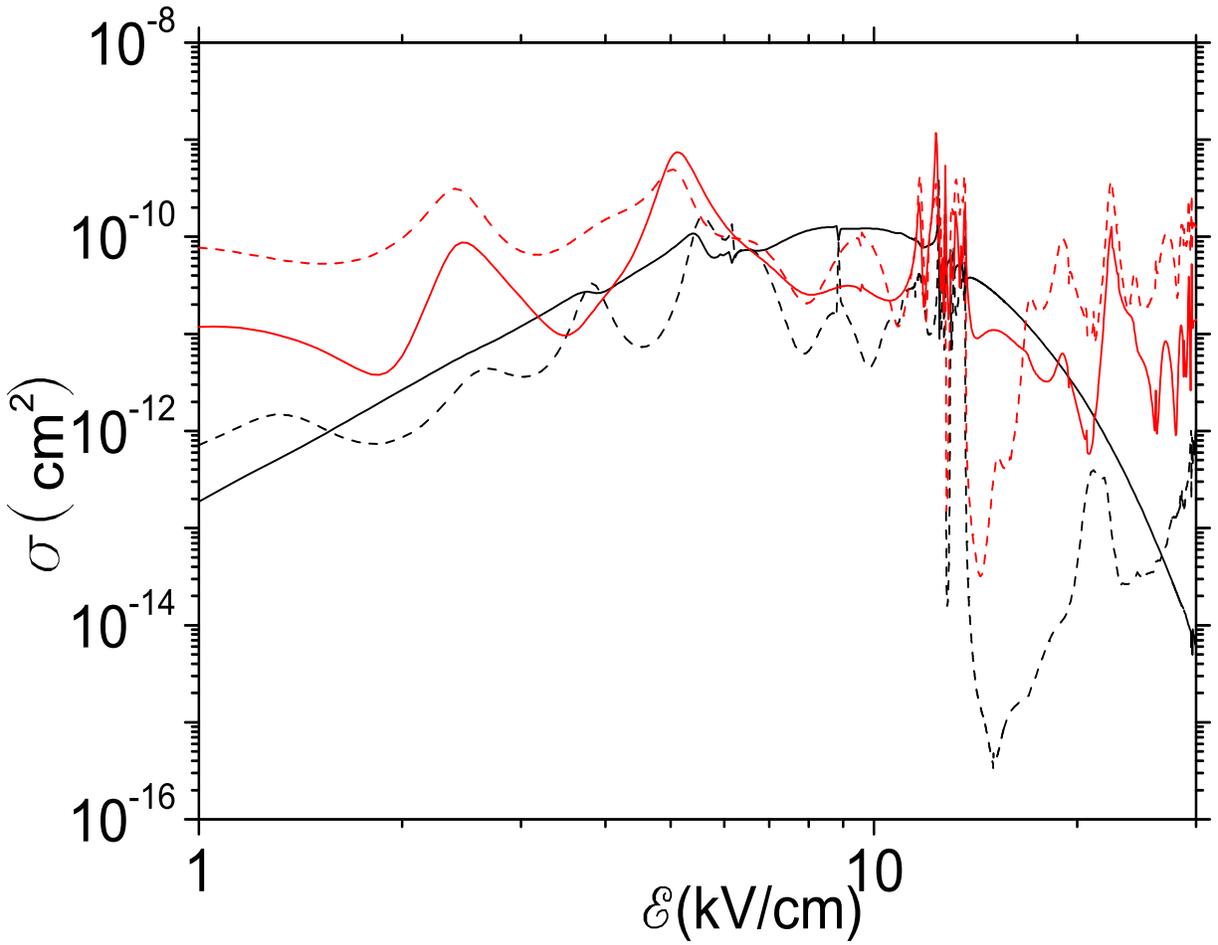}}
\caption{ Elastic(solid lines) and inelastic (dashed lines) cross
sections versus electrostatic field at collision energy $1\mu K$
for ClCN molecule.Thick and thin curves are for Fermi and Bose
particles respectively.
 $M_{tot}=0$}.
\label{e_bf}
\end{figure}

\begin{figure}
\centerline{\includegraphics[width=1.08\linewidth,height=0.9\linewidth,angle=0]{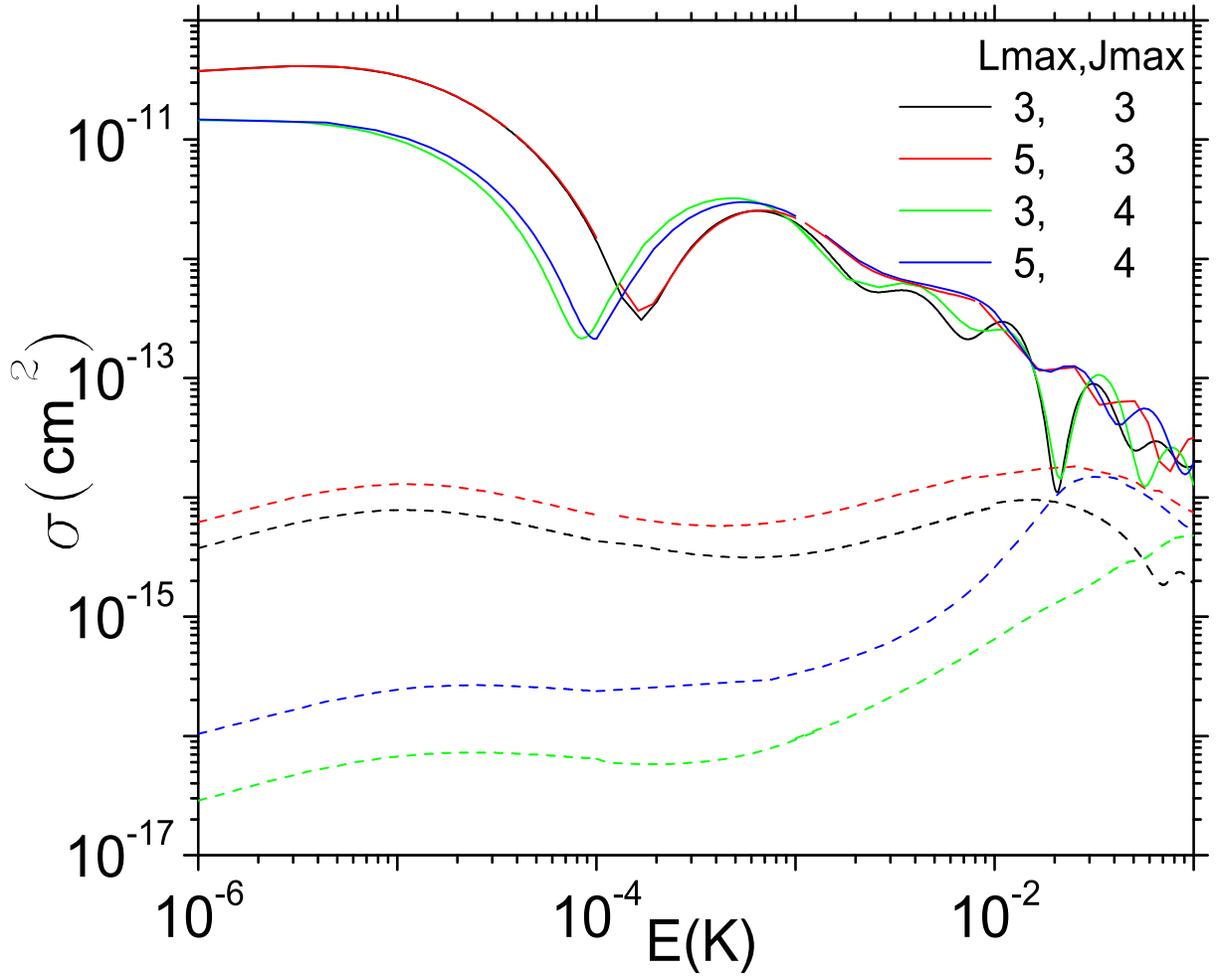}}
\caption{Elastic(solid lines) and inelastic (dashed lines) cross
sections versus the collisional energy for ClCN molecule. $J_{max}$
and $L_{max}$ are maximal taken into account values of rotation an
partial quantum numbers. The electrostatic field is $14 kV/cm$}
\label{converg}
\end{figure}

\end{document}